\documentclass[11pt]{llncs}
\usepackage{fullpage}
\usepackage{graphicx}

\newcommand{\Oh}[1]
    {\ensuremath{\mathcal{O}\!\left( {#1} \right)}}
\newcommand{\bwt}
    {\ensuremath{\mathrm{bwt}}}
\newcommand{\occ}
    {\ensuremath{\mathrm{occ}}}
\newcommand{\locate}
    {\ensuremath{\mathrm{locate}}}
\newcommand{\antilocate}
    {\ensuremath{\mbox{anti-locate}}}

\begin{document}

\title{Multi-Pattern Matching in a\\Compressed Distributed Index}
\author{Travis Gagie\inst{1}
    \and Kalle Karhu\inst{1}
    \and Juha K\"arkk\"ainen\inst{2}
    \and\\ Veli M\"akinen\inst{2}
    \and Leena Salmela\inst{2}\thanks{Supported by Academy of Finland grant 118653 (\mbox{ALGODAN})}
    \and Jorma Tarhio\inst{1}}
\institute{Aalto University, Finland\\
    \email{\{travis.gagie,\,kalle.karhu,\,jorma.tarhio\}@aalto.fi}\\
    \mbox{}\\
    University of Helsinki, Finland\\
    \email{\{juha.karkkainen,\,veli.makinen,\,leena.salmela\}@cs.helsinki.fi}}
\maketitle

\begin{abstract}
Compressed full-text indexes have been one of pattern matching's most important success stories of the past decade.  We can now store a text in nearly the information-theoretic minimum of space, such that we can still quickly count and locate occurrences of any given pattern.  However, some files or collections of files are so huge that, even compressed, they do not all fit in one machine's internal memory.  One solution is to break the file or collection into pieces and create a distributed index spread across many machines (e.g., a cluster, grid or cloud).  Suppose we want to search such an index for many patterns.  Since each pattern is to be sought on each machine, it is worth spending a reasonable amount of time to preprocess the patterns if that leads to faster searches.  In this paper we show that if the concatenation of the patterns can be compressed well with LZ77, then we can take advantage of their similarities to speed up searches in BWT-based indexes.  More specifically, if we are searching for $t$ patterns of total length $m$ in a distributed index for a text of total length $n$, then we spend $\Oh{m + (g + t) \log m}$ time preprocessing the patterns on one machine and then $\Oh{(g + t) \log^2 m \log^{1 + \epsilon} n}$ time searching for them on each machine, where $g$ is the size of the smallest straight-line program for the concatenation of the patterns.  Thus, if the concatenation of the patterns has a small straight-line program --- plausible if the patterns are similar --- and the number of machines is large, we achieve a theoretically significant speed-up.  The techniques we use seem likely to be of independent interest and we show how they can be applied to pattern matching with wildcards and parallel pattern matching.
\end{abstract}

\section{Introduction} \label{sec:introduction}

Compressed full-text indexes have revolutionized some areas of pattern matching, offering both nearly optimal compression and fast searching simultaneously, but other areas have yet to benefit from them.  For example, when Navarro and M\"akinen~\cite{NM07} wrote their survey of such indexes, most of the literature on them dealt with exact, single-pattern matching with one processor, with a few notable papers dealing with approximate matching.  Since then, research on those topics has continued and research has begun on, e.g., matching with wildcards~\cite{LSTY07}, parallelized searching and distributed indexes~\cite{RNO10}. As far as we know, however, there has been no previous work on designing indexes for multi-pattern matching in the sense of, say, the Aho-Corasick algorithm~\cite{AC75}.  That is, although indexing really makes sense only when searching for multiple patterns --- if we are to search only for one, then it is faster to do so directly than to first build an index --- the standard approach is to search for the patterns separately (see, e.g.,~\cite[pg. 1]{ALLS07}) without taking advantage of possible similarities between them.  In this paper we show that, if the concatenation of the patterns can be compressed well with LZ77~\cite{ZL77}, then we can take advantage of their similarities to speed up searches in indexes based on the Burrows-Wheeler Transform (BWT)~\cite{BW94}.  Since running LZ can be as time-consuming as searching in the index directly, we consider the case when we are searching for many patterns in an index for a file so large that, even compressed, it does not all fit in one machine's memory and must be stored as a distributed index spread across many machines (e.g., a cluster, grid or cloud).  Since each pattern is to be sought on each machine, it is worth spending a reasonable amount of time to preprocess the patterns if that leads to faster searches.  More specifically, if we are searching for $t$ patterns of total length $m$ in a distributed index for a text of total length $n$, then we spend $\Oh{m + (g + t) \log m}$ time preprocessing the patterns on one machine and then $\Oh{(g + t) \log^2 m \log^{1 + \epsilon} n}$ time searching for them on each machine, where $g$ is the size of the smallest straight-line program for the concatenation of the patterns.  Thus, if the concatenation of the patterns has a small straight-line program --- plausible if the patterns are similar --- and the number of machines is large, we achieve a theoretically significant speed-up.  Such a speed-up could be of practical importance in several bioinformatics applications, in which both very large files and multi-pattern matching are common~\cite{MA11,RSKKT09}.

\begin{figure}[t]
\begin{center}
\resizebox{30ex}{!}{\includegraphics{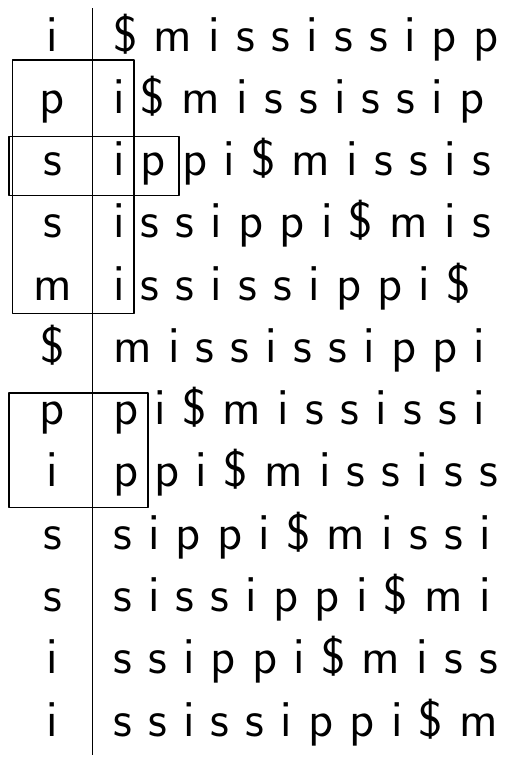}}
\caption{The BWT of $\mathsf{mississippi}$ and the intervals \([2, 5]\), \([7, 8]\) and \([3]\) corresponding to $\mathsf{i}$, $\mathsf{p}$ and $\mathsf{ip}$, respectively.}
\end{center}
\label{fig:bwt}
\end{figure}

The BWT sorts the characters in a text $T$ (possibly with a special end-of-string character \$ appended) into the lexicographic order of the suffixes immediately following them.  As a result, any pattern has a corresponding interval in the BWT, containing the character immediately before each occurrence of that pattern.  For example, if \(T = \mathsf{mississippi}\), then \(\bwt (T) = \mathsf{ipssm\$pissii}\); the intervals corresponding to patterns $\mathsf{i}$, $\mathsf{p}$ and $\mathsf{ip}$ are \([2, 5]\), \([7, 8]\) and \([3]\), respectively.  This is illustrated in Figure~\ref{fig:bwt}.

BWT-based indexes are among the most competitive compressed full-text indexes known.  They can store a text in nearly the information-theoretic minimum of space while still allowing us to quickly count and locate occurrences of any given pattern.  For more information on the BWT and on such indexes in general, we refer the reader to the recent book by Adjeroh, Bell and Mukherjee~\cite{ABM08}.  Barbay, Gagie, Navarro and Nekrich~\cite{BGNN10} very recently gave one such index that stores a text \(T [1..n]\) over an alphabet of size $\sigma$ in \(n H_k (T) + o (n) (H_k (T) + 1)\) bits, for all \(k \leq (1 - \epsilon) \log_\sigma n - 1\) simultaneously, such that:
\begin{itemize}
\item given a pattern \(P [1..m]\), in $\Oh{m \log \log \sigma}$ time we can find the endpoints of the interval in \(\bwt (T)\) containing the character immediately before each occurrence of $P$ in $T$ (and, thus, count the number of occurrences by taking their difference and adding 1);
\item given a character \(T [i]\)'s position in \(\bwt (T)\), in $\Oh{\log^{1 + \epsilon} n}$ time we can find $i$.
\end{itemize}
The second query is called $\locate$ and will, together with the inverse of it that we define in Section~\ref{sec:anti-locate}, be useful to us not just as a query to be implemented for its own sake, as is usual, but also as a primitive to implement other queries.

Our idea is to take advantage of long repeated sub-patterns: after all, if we have already spent the time searching for a sub-pattern, then we would like to avoid searching for it again.  To quantify our advantage, our analyses are in terms of the number of phrases in the LZ77 parse of the concatenation of the patterns, and the size of the smallest straight-line program that generates the concatenation.  To find common sub-patterns, we compute the LZ77 parse of the concatenation of the patterns.  We consider the version of LZ77 that requires the match to be completely contained in the prefix already parsed.  The parse is defined in terms of a greedy algorithm: if we are parsing a pattern \(P [1..m]\) and have already processed \(P [1..i]\), then we look for the longest prefix of \(P [i + 1..m]\) that we have already seen; we record the position and length of the matching sub-pattern, or \(P [i + 1]\) if it does not exist. Rytter~\cite{Ryt03} showed that the number of phrases in this LZ77 parse for a string is a lower bound on the size of the smallest context-free grammar in Chomsky normal form (or straight-line program) that generates that string and only that string.  He also showed how to convert the LZ77 parse into a straight-line program with a logarithmic blow-up.

In Section~\ref{sec:anti-locate} we define the $\antilocate$ query, show how it can be implemented by adding \(o (n)\) bits to Barbay et al.'s index and, as a warm-up, show how it is useful in pattern matching with wildcards.  In Section~\ref{sec:concatenating} we show how we can use $\locate$ and $\antilocate$ to find the interval in \(\bwt (T)\) for the concatenation of two sub-patterns in polylogarithmic time, assuming we already know the intervals for those sub-patterns.  This means that, if we have a straight-line program for a pattern, then we can find the interval for that pattern using polylogarithmic time per distinct non-terminal in the program.  From this we obtain our speed-up for multi-pattern matching in a distributed index.  In Section~\ref{sec:parallel} we show how we can parallelize the searching, obtaining a speed-up linear in the number of processors.  In Section~\ref{sec:applications} we discuss some other possible applications, and we summarize our results in Section~\ref{sec:conclusions}.

\section{Anti-Locate and Pattern Matching with Wildcards} \label{sec:anti-locate}

Like many other BWT-based indexes, Barbay et al.'s~\cite{BGNN10} uses a \(o (n)\)-bit sample to support $\locate$, which takes the position of a character in \(\bwt (T)\) and returns that character's position in $T$.  We can store a similar sample for the inverse query: $\antilocate$ takes the position of a character in $T$ and returns that character's position in \(\bwt (T)\).  We store the position in \(\bwt (T)\) of every \((\log n \log \log n)\)th character of $T$; given $i$, in $\Oh{1}$ time we find the character whose position in \(\bwt (T)\) we have stored and that is closest to \(T [i]\) in $T$; then from that we use rank and select queries to find \(T [i]\)'s position in \(\bwt (T)\) in \(\Oh{\log n \log \log n \log \log \sigma} \subset \Oh{\log^{1 + \epsilon} n}\) time.

\begin{lemma} \label{lem:anti-locate}
We can add \(o (n)\) bits to Barbay et al.'s index such that it supports $\antilocate$ in $\Oh{\log^{1 + \epsilon} n}$ time.
\end{lemma}

To give a simple illustration of how $\antilocate$ can be useful, in the rest of this section we apply it to pattern matching with wildcards.  Lam, Sung, Tam and Yiu~\cite{LSTY07} gave an $\Oh{n \log n}$-bit index for a text \(T [1..n]\) such that, given a pattern \(P [1..m] = P_1 ?^{w_1} P_2 ?^{w_2} \ldots P_t\) containing a total of $w$ occurrences of the wildcard symbol $?$ and $t$ maximal sub-patterns \(P_1, \ldots, P_t\) containing no wildcards, we can find all substrings of $T$ matching $P$ in $\Oh{m + t \min_h \{\occ (P_h)\}}$ time when each wildcard must be replaced by a character; when wildcards can be replaced or ignored, we use $\Oh{m + w t \min_h \{\occ (P_h)\}}$ time.  (Notice we can ignore wildcards at the beginning or end of $P$.)  We give the first compressed index for pattern matching with wildcards, by showing how we can search our index from Lemma~\ref{lem:anti-locate} in $\Oh{m \log \log \sigma + t \min_h \{\occ (P_h)\} \log^{1 + \epsilon} n}$ time when wildcards must be replaced, or $\Oh{m \log \log \sigma + w t \min_h \{\occ (P_h)\} \log^{1 + \epsilon} n}$ time when they can be replaced or ignored.

First assume that each wildcard must be replaced by a character.  We first find the intervals in $\bwt (T)$ corresponding to each of \(P_1, \ldots, P_t\), which takes $\Oh{m \log \log \sigma}$ time.  We choose the shortest such interval, which has length $\min_h \{\occ (P_h)\}$; suppose it is for sub-pattern $P_j$.  If \(j \leq t\), we check each position $i$ in the interval for $P_j$ to see whether
\[\antilocate (\locate (i) + |P_j| + w_j)\]
is in the interval for $P_{j + 1}$; if so, then we have found the starting position in \(\bwt (T)\) of a substring in $T$ matching \(P_j ?^{w_j} P_{j + 1}\).  If \(j = t\), we check each position $i$ in the interval for $P_j$ to see whether
\[\antilocate (\locate (i) - w_{j - 1} - |P_{j - 1}|)\]
is in the interval for $P_{j - 1}$; if so, then we have found the starting position in \(\bwt (T)\) of a substring in $T$ matching \(P_{j - 1} ?^{w_{j - 1}} P_j\).  In either case, this takes $\Oh{\occ (P_j) \log^{1 + \epsilon} n}$ time and yields the positions in \(\bwt (T)\) for at most \(\occ (P_j)\) matching substrings.  For example, suppose we want to match {\sf s??s} in \(T = \mathsf{mississippi}\).  The interval for {\sf s} is \([9, 12]\) and
\begin{eqnarray*}
\antilocate (\locate (9) + |\mathsf{s}| + 2) = \antilocate (9) & = & 7\\
\antilocate (\locate (10) + |\mathsf{s}| + 2) = \antilocate (6) & = & 9\\
\antilocate (\locate (11) + |\mathsf{s}| + 2) = \antilocate (8) & = & 8\\
\antilocate (\locate (12) + |\mathsf{s}| + 2) = \antilocate (5) & = & 11\,,
\end{eqnarray*}
so the intervals in the intersection, \([9]\) and \([11]\), correspond to the two substrings, {\sf siss} and {\sf ssis}, that match {\sf s??s}.

Because of the wildcards, the positions we have found may not be consecutive.  Nevertheless, we can repeat the procedure above for each of them, to append or prepend sequences of wildcards and sub-patterns, again using a total of $\Oh{\occ (P_j) \log^{1 + \epsilon} n}$ time for each sub-pattern.  It follows that we can find all the substrings of $T$ matching $P$ in $\Oh{m \log \log \sigma + t\min_h \{\occ (P_h)\} \log^{1 + \epsilon} n}$ time.

Now assume wildcards can be replaced or ignored.  We proceed much as before but whenever we would check \(\antilocate (\locate (i) + |P_j| + w_j)\) or \(\antilocate (\locate (i) - w_j - |P_j|)\) for some $i$ and $j$, we now check \(\antilocate (\locate (i) + |P_j| + w_j')\) or \(\antilocate (\locate (i) - w_j' - |P_j|)\) for \(0 \leq w_j' \leq w_j\).  Calculation shows that the whole procedure now takes $\Oh{m \log \log \sigma + w t \min_h \{\occ (P_h)\} \log^{1 + \epsilon} n}$ time.

\begin{theorem} \label{thm:wildcards}
We can build an \((n H_k (T) + o (n (H_k (T) + 1)))\)-bit index for a text \(T [1..n]\) such that, given a pattern \(P [1..m]\) containing $w$ wildcards and $t$ maximal sub-patterns \(P_1, \ldots, P_t\) containing no wildcards, we can find all substrings of $T$ matching $P$ in $\Oh{m \log \log \sigma + t \min_h \{\occ (P_h)\} \log^{1 + \epsilon} n}$ time when wildcards must be replaced, or $\Oh{m \log \log \sigma + w t \min_h \{\occ (P_h)\} \log^{1 + \epsilon} n}$ time when they can be replaced or ignored.
\end{theorem}

\section{Concatenating Sub-Patterns} \label{sec:concatenating}

\begin{figure}[t]
\begin{center}
\resizebox{100ex}{!}{\includegraphics{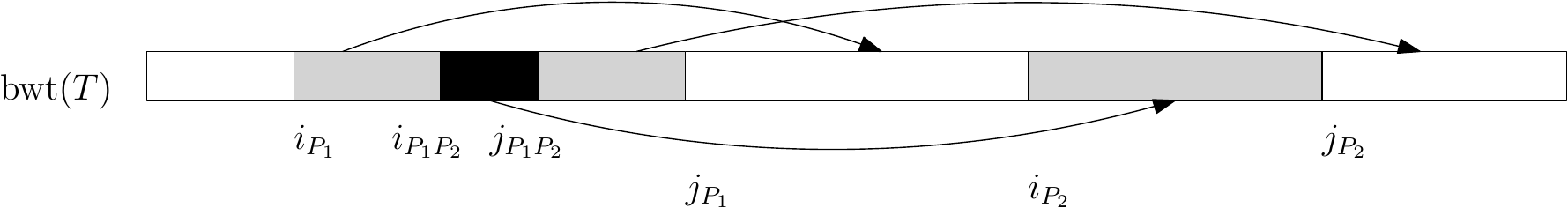}}
\caption[]{\label{fig:intervals}The BWT of $T$ with the intervals \([i_{P_1}, j_{P_1}]\) and \([i_{P_2}, j_{P_2}]\) corresponding to $P_1$ and $P_2$ shown in grey and the interval \([i_{P_1 P_2}, j_{P_1 P_2}]\) corresponding to \(P_1 P_2\) shown in black.  The three cases shown are:
\protect\linebreak[4]
\mbox{
\begin{minipage}[t]{0.95\textwidth}
\begin{itemize}
\item if \(i_{P_1} \leq i < i_{P_1 P_2}\) then \(\antilocate (\locate (i) + |P_1|) < i_{P_2}\) (left arrow);
\item if \(i_{P_1 P_2} \leq i \leq j_{P_1 P_2}\) then \(i_{P_2} \leq \antilocate (\locate (i) + |P_1|) \leq j_{P_2}\) (center arrow);
\item if \(j_{P_1 P_2} < i \leq j_{P_1}\) then \(\antilocate (\locate (i) + |P_1|) > j_{P_2}\) (right arrow).
\end{itemize}
\end{minipage}}}
\end{center}
\end{figure}

We now describe the key observation behind our main result: how we can use $\antilocate$ when we have found the intervals in \(\bwt (T)\) corresponding to sub-patterns $P_1$ and $P_2$ and now want to find the interval corresponding to \(P_1 P_2\).  Let $i_{P_1}$ and $j_{P_1}$ be the endpoints of the interval for $P_1$, let $i_{P_2}$ and $j_{P_2}$ be the endpoints of the interval for $P_2$, and let $i_{P_1 P_2}$ and $j_{P_1 P_2}$ be the endpoints of the interval for \(P_1 P_2\) as shown in Figure~\ref{fig:intervals}.

Notice that, since every occurrence of \(P_1 P_2\) in $T$ is also an occurrence of $P_1$, \([i_{P_1 P_2}, j_{P_1 P_2}]\) is a subinterval of \([i_{P_1}, j_{P_1}]\).  Also, if $i$ is in \([i_{P_1}, j_{P_1}]\) but \(\antilocate (\locate (i) + |P_1|)\) is strictly before $i_{P_2}$ then, by the definition of the Burrows-Wheeler Transform, \(T [\locate (i) + |P_1| + 1..\locate (i) + |P_1| + |P_2|]\) is lexicographically strictly less than $P_2$, so \(T [\locate (i) + 1..\locate (i) + |P_1| + |P_2|]\) is lexicographically strictly less than \(P_1 P_2\) and $i$ is strictly before $i_{P_1 P_2}$.  (Recall that the characters in the interval \([i_P, j_P]\) for a pattern $P$ occur in $T$ immediately before occurrences of $P$; therefore, if $i$ is in \([i_P, j_P]\), then the corresponding occurrence of $P$ is \(T [\locate (i) + 1..\locate (i) + |P|]\), rather than \(T [\locate (i)..\locate (i) + |P| - 1]\).)  If \(\antilocate (\locate (i) + |P_1|)\) is in \([i_{P_2}, j_{P_2}]\), then \(T [\locate (i) + |P_1| + 1..\locate (i) + |P_1| + |P_2|]\) is an occurrence of $P_2$, so \(T [\locate (i) + 1..\locate (i) + |P_1| + |P_2|]\) is an occurrence of \(P_1 P_2\) and $i$ is in \([i_{P_1 P_2}, j_{P_1 P_2}]\).  Finally, if \(\antilocate (\locate (i) + |P_1|)\) is strictly after $j_{P_2}$, then \(T [\locate (i) + |P_1| + 1..\locate (i) + |P_1| + |P_2|]\) is lexicographically strictly greater than $P_2$, so \(T [\locate (i) + 1..\locate (i) + |P_1| + |P_2|]\) is lexicographically strictly greater than \(P_1 P_2\) and $i$ is strictly after $j_{P_1 P_2}$.  Figure~\ref{fig:intervals} illustrates these three cases.  It follows that, by using binary search in \([i_{P_1}, j_{P_1}]\), we can find $i_{P_1 P_2}$ and $j_{P_1 P_2}$ in \(\Oh{\log (j_{P_1} - i_{P_1}) \log^{1 + \epsilon} n} = \Oh{\log \occ (P_1) \log^{1 + \epsilon} n}\) time.  For example, suppose \(T = \mathsf{mississippi}\) and we have found the intervals \([2, 5]\) and \([7, 8]\) for {\sf i} and {\sf p}, respectively (shown in Figure~\ref{fig:bwt}), and now want to find the interval for {\sf ip}. A binary search through the values
\begin{eqnarray*}
\antilocate(\locate (2) + |\mathsf{i}|) = \antilocate (11) & = & 1\\
\antilocate(\locate (3) + |\mathsf{i}|) = \antilocate (8) & = & 8\\
\antilocate(\locate (4) + |\mathsf{i}|) = \antilocate (5) & = & 11\\
\antilocate(\locate (5) + |\mathsf{i}|) = \antilocate (2) & = & 12
\end{eqnarray*}
shows that only 8 is in \([7, 8]\), so the interval for {\sf ip} is \([3]\).  We show all four of the values above for the sake of exposition, even though  a binary search requires us to evaluate only some of them.

We can improve this by sampling the value \(\antilocate (\locate (i) + \ell)\) for every \((\ell \log^{2 + \epsilon} n)\)th position $i$ in \(\bwt (T)\), for \(1 \leq \ell \leq n\), which takes \(\Oh{\sum_\ell \frac{n \log n}{\ell \log^{2 + \epsilon} n}} = o (n)\) bits.  We can now find $i_{P_1 P_2}$ and $j_{P_1 P_2}$ as follows: we first use binary search in the values sampled for \(\ell = |P_1|\) in the interval \([i_{P_1}, j_{P_1}]\) of \(\bwt (T)\) to find subintervals of length at most \(|P_1| \log^{2 + \epsilon} n\) that contain $i_{P_1 P_2}$ and $j_{P_1 P_2}$, which takes $\Oh{\log n}$ time since we do not need to perform $\locate$ and $\antilocate$ queries here; we then use binary search in those subintervals to find $i_{P_1 P_2}$ and $j_{P_1 P_2}$, which takes $\Oh{\log |P_1| \log^{1 + \epsilon} n \log \log n}$ time.  The \(\log \log n\) factor can be hidden within the \(\log^{1 + \epsilon} n\) factor.

\begin{lemma} \label{lem:gluing}
We can add \(o (n)\) bits to Barbay et al.'s index such that, once we have found the intervals in \(\bwt (T)\) corresponding to $P_1$ and $P_2$, we can find the interval corresponding to \(P_1 P_2\) in $\Oh{\log |P_1| \log^{1 + \epsilon} n}$ time.
\end{lemma}

Let \(P [1..m]\) be a pattern.  Notice that, if we have a context-free grammar in Chomsky normal form that generates $P$ and only $P$, also known as a straight-line program (SLP) for $P$, then we can find the interval in \(\bwt (T)\) corresponding to $P$ by applying Lemma~\ref{lem:gluing} once for each distinct non-terminal $X$: assuming we have already found the intervals for the expansions of the symbols on the right-hand side of the unique rule in which $X$ appears on the left, Lemma~\ref{lem:gluing} yields the interval for the expansion of $X$.  For example, for the SLP
\vspace{5ex}
\newline
\begin{tabular}{lr}
\parbox{30ex}
{\vspace{-25ex}
\begin{eqnarray*}
X_7 & \rightarrow & X_6 X_5\\
X_6 & \rightarrow & X_5 X_4\\
X_5 & \rightarrow & X_4 X_3\\
X_4 & \rightarrow & X_3 X_2\\
X_3 & \rightarrow & X_2 X_1\\
X_2 & \rightarrow & a\\
X_1 & \rightarrow & b
\end{eqnarray*}}
&
\resizebox{60ex}{!}{\includegraphics{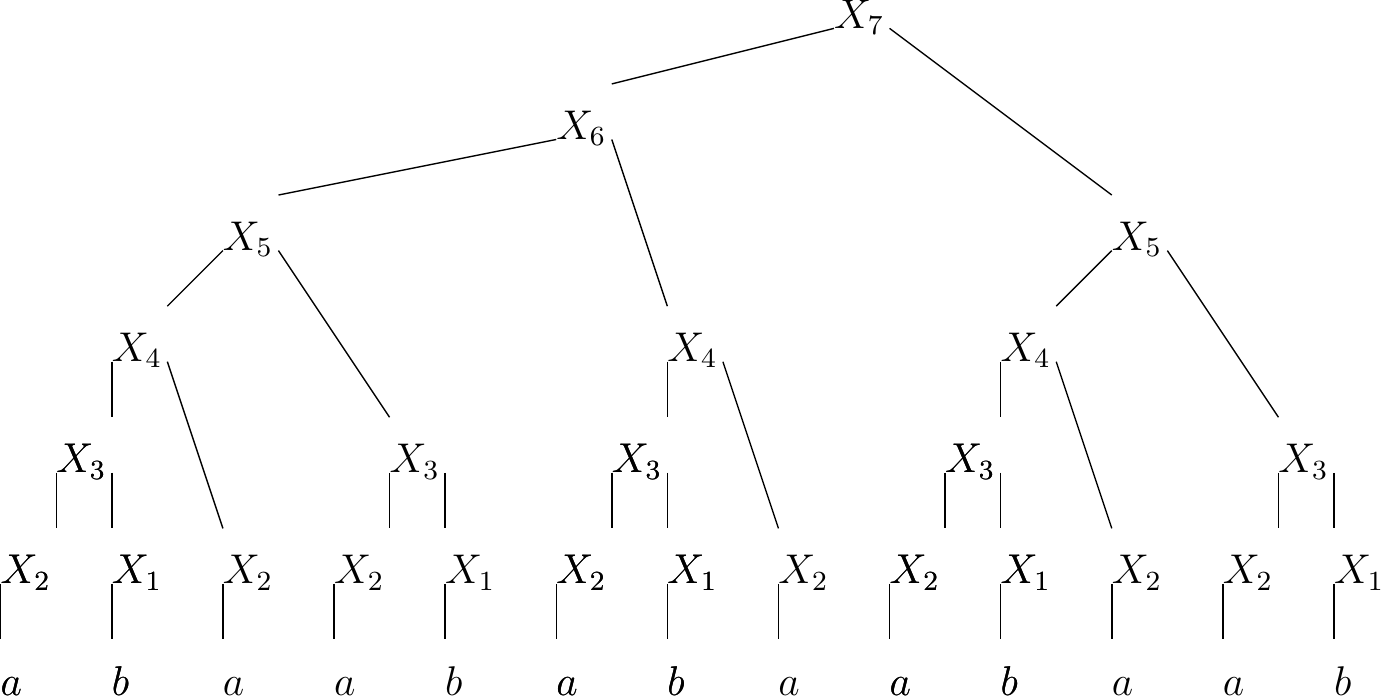}}
\end{tabular}
\vspace{5ex}
\newline
which generates \(a b a a b a b a a b a a b\), we perform searches for $a$ and $b$ to find the intervals for the (single-character) expansions $X_1$ and $X_2$ and apply Lemma~\ref{lem:gluing} to find, in turn, the intervals for the expansions of \(X_3, \ldots, X_7\).  This is like working from the leaves to the root of the parse-tree (shown above on the right), but we note we need apply Lemma~\ref{lem:gluing} only once for each distinct non-terminal, rather than for every node of the tree.

Rytter~\cite{Ryt03} gave an algorithm for building an SLP with nearly minimum size.  He first proved that the number of phrases in the LZ77 parse of $P$, even without allowing overlaps, is a lower bound on the size of any SLP.  He then showed how to convert that parse into an SLP.  To do this, he defined an AVL-grammar for $P$ to be a context-free grammar in Chomsky normal form that generates $P$ and only $P$, such that the parse-tree has the shape of an AVL-tree.  Notice this means the parse-tree has height $\Oh{\log m}$, a fact we will use in Section~\ref{sec:parallel}.

Suppose the first $i$ phrases of the LZ77 parse encode \(P [1..j]\), we have already built an AVL-grammar for \(P [1..j]\) and the \((i + 1)\)st phrase is \(\langle b, \ell \rangle\).  Then we can build the AVL-grammar for \(P [1..j + \ell]\) by splitting the parse-tree (as an AVL-tree) for \(P [1..j]\) between its \((b - 1)\)st and $b$th leaves and between its \((b + \ell - 1)\)st and \((b + \ell)\)th leaves, so as to obtain an AVL-grammar for \(P [b..b + \ell + 1]\), then joining that to the right side of the AVL-grammar for \(P [1..j]\).  Rytter showed how to do this in $\Oh{\log j}$ time while adding $\Oh{\log j}$ new non-terminals.  If we repeat this procedure for each phrase, then in $\Oh{g \log m}$ time we obtain an SLP with $\Oh{g \log m}$ non-terminals, where $g$ is the size of the smallest SLP.

If we apply LZ77 parsing to a sequence of patterns \(P_1, \ldots, P_t\) one by one, while allowing matches to cross the boundaries of previously seen patterns, then we produce at most $t$ more phrases than if we processed the concatenation of the patterns \(P_1 \ldots P_t\) as a single string.  To see why, consider the parse for the concatenation \(P_1 \ldots P_t\).  If any phrase crosses the boundary between patterns, then we break the phrase at the boundary.  This increases the number of phrases by at most $t$.  Therefore, since LZ77's greedy parsing is optimal, if follows that parsing \(P_1, \ldots, P_t\) separately produces at most $t$ more phrases than parsing them concatenated.  Therefore, Rytter's algorithm on such a parse produces an AVL-grammar with size $\Oh{(g + t) \log m}$.  Once we have such a grammar for \(P_1 \ldots P_t\), we can perform \(t - 1\) splits in $\Oh{t \log m}$ time to obtain grammars for each of \(P_1, \ldots, P_t\), adding $\Oh{t \log m}$ new non-terminals.  Applying Lemma~\ref{lem:gluing} to each non-terminal in these separate AVL-grammars, we obtain the following theorem.

\begin{theorem} \label{thm:grammar}
We can build an \((n H_k (T) + o (n (H_k (T) + 1)))\)-bit index for a text \(T [1..n]\) such that, if we are given Rytter's AVL-grammar for the concatenation of $t$ patterns \(P_1, \ldots, P_t\) of total length $m$, parsed one by one, then we can search for \(P_1, \ldots, P_t\) in $\Oh{(g + t) \log^2 m \log^{1 + \epsilon} n}$ time, where $g$ is the size of the smallest SLP for the concatenation \(P_1 \ldots P_t\).
\end{theorem}

Of course, since computing the LZ77 parse takes linear time~\cite{CPS08}, Theorem~\ref{thm:grammar} does not help us much when searching in only one index --- we can achieve at best a factor of \(\log \log \sigma\) speed-up over searching directly.  If we compute Rytter's AVL-grammar on one machine and then send the resulting SLP to $q$ machines each storing part of a distributed index, however, then we use $\Oh{m + (g + t) \log m}$ preprocessing time on the first machine but then only $\Oh{(g + t) \log^2 m \log^{1 + \epsilon} n}$ time on each machine, rather than $\Oh{m \log \log \sigma}$ on each machine.  This is our main result for this paper, although we feel Theorem~\ref{thm:grammar} and the techniques behind it are likely to prove of independent interest.

\begin{corollary} \label{cor:distributed}
Given texts \(T_1, \ldots, T_q\) to be stored on $q$ machines, we can build indexes of size \(|T_1| H_k (T_1) + o (|T_1| (H_k (T_1) + 1)), \ldots, |T_q| H_k (T_q) + o (|T_q| (H_k (T_q) + 1))\) bits such that, given $t$ patterns \(P_1, \ldots, P_t\) of total length $m$, we can spend $\Oh{m + (g + t) \log m}$ time preprocessing \(P_1, \ldots, P_t\) on one machine, where $g$ is the size of the smallest SLP for the concatenation \(P_1 \ldots P_t\), then $\Oh{(g + t) \log^2 m \log^{1 + \epsilon} n}$ time searching on each machine.
\end{corollary}

In other words, when the patterns are compressible together (e.g., when they have small edit distance, or when most can be formed by cutting and pasting parts of the others) then we can greatly reduce the total amount of processing needed to search in a distributed index.

\section{Parallel Searching} \label{sec:parallel}

Several authors (see~\cite{KW05} and references therein) have already studied parallelization of LZ77 or its variants, so in this section we focus on parallelizing the actual searches.  Suppose we are searching for a pattern \(P [1..m]\) in a text \(T [1..n]\) on a machine with $p$ processors.  Russo, Navarro and Oliveira~\cite{RNO10} gave an \((n H_k (T) + o (n \log \sigma))\)-bit index for $T$ that we can search in
\[\Oh{m / p + \log n \log \log n (\log p + \log \log n \log \log p)}\]
time.  For reasonably long patterns and not too many processors, their speedup is linear in $p$.  We now show that if the smallest SLP for $P$ has size $g$ and we already have the AVL-grammar for $P$ that results from Rytter's algorithm, then we can search our index from Lemma~\ref{lem:gluing} in $\Oh{\lceil g / p \rceil \log^2 m \log^{1 + \epsilon} n}$ time.  That is, we achieve an unconditionally linear speedup over Theorem~\ref{thm:grammar} and a better upper bound than Russo et al. when $P$ is very compressible.

Each non-terminal in the AVL-grammar can appear only at a specific $\Oh{\log m}$ height in the parse tree for $P$.  We sort the non-terminals into non-decreasing order by height and process them in that order.  Notice that the concatenation we perform for each non-terminal cannot depend on the concatenation for any non-terminal of equal or greater height.  Therefore, we can parallelize the concatenations for non-terminals of the same height.  For each height with $r$ non-terminals, we use $\Oh{\lceil r / p \rceil \log m \log^{1 + \epsilon} n}$ time.  Since there are $\Oh{g \log m}$ non-terminals in total and $\Oh{\log m}$ possible heights, calculation shows we use $\Oh{\lceil g / p \rceil \log^2 m \log^{1 + \epsilon} n}$ time.

\begin{theorem} \label{thm:parallel}
We can build an \((n H_k (T) + o (n (H_k (T) + 1)))\)-bit index for a text \(T [1..n]\) such that, on a machine with $p$ processors and given Rytter's AVL-grammar for a pattern \(P [1..m]\) whose smallest SLP has size $g$, we can search for $P$ in $\Oh{\lceil g / p \rceil \log^2 m \log^{1 + \epsilon} n}$ time.
\end{theorem}

\section{Other Applications} \label{sec:applications}

There are several other possible applications for the techniques we have developed in this paper.  For example, we might be given patterns with wildcards to preprocess, where the characters that will replace those wildcards will be given to us later.  Such a pattern could be a fragment of DNA with wildcards in the locations of single-nucleotide polymorphisms, which we can use as a re-useable template: we search in advance for the maximal sub-patterns not containing any wildcards so that later, given an assignment of characters to the wildcards, we can quickly find all occurrences of the filled-in pattern.  We can even allow the wildcards to represent blanks of unknown length, so the template could be a part of the document with missing characters, words or phrases.

In Section~\ref{sec:concatenating} we preprocess \(P_1, \ldots, P_t\) first, send the resulting grammars to each machine, then use those grammars and Lemma~\ref{lem:gluing} to speed up searches in each part of the distributed index.  Since applying Lemma~\ref{lem:gluing} on a specific machine yields an interval in the part of the distributed index stored on that machine, in this context it makes little sense to mix the preprocessing and the searching.  In general, however, we can apply Lemma~\ref{lem:gluing} to each non-terminal as it is created; we can also split off the SLP for each pattern when we have finished parsing that pattern.  This could be useful if, for example, we want to search for the patterns as they are given to us, instead of batching them, and we are given the LZ77 parse instead of having to compute it ourselves.  More generally, we could be given any set of instructions on how to form each new pattern by cutting and pasting parts of the patterns we have already seen.

We can extend this idea to consider maintaining dynamic libraries of sub-patterns: we keep an AVL-grammar for each sub-pattern, with the interval for each non-terminal stored; given instructions on how to assemble a new pattern by cutting and pasting pieces of the sub-patterns, we (non-destructively) split the AVL-grammars for those sub-patterns and form the new pattern, simultaneously computing the interval for the new pattern.  We leave as future work exploring these and other possible applications.  We are currently investigating whether our results can be used to speed up approximate pattern matching in compressed indexes (see~\cite{RNOM09} for a recent discussion of this topic).  Notice each pair of strings within a small edit distance of a pattern share long substrings; we can use our results and heuristics to explore adaptively the neighborhood around the pattern, pruning branches of our search once we know they cannot yield a match.

In some of these applications, of course, Barbay et al.'s index may not be the best choice.  We chose it for this paper because it has the smallest space bound but, when time is more important than space, we could use, e.g., Sadakane's Compressed Suffix Array~\cite{Sad03}: this index takes \(\frac{1 + \delta_1}{\delta_2} n H_0 (s) + 2 n \log (H_0 (s) + 1) + 3 n + o (n)\) bits, where $\delta_1$ and \(\delta_2 < 1\) are arbitrary positive constants, and supports both $\locate$ and (as we will show in the full version of this paper) $\antilocate$ in $\Oh{\log^{\delta_2} n / (\delta_1 \delta_2)}$ time without any additional data structures, rather than the $\Oh{\log^{1 + \epsilon} n}$ time we used in this paper.

\section{Conclusions} \label{sec:conclusions}

We have shown how, if we have already found the intervals in the BWT corresponding to two patterns, then in polylogarithmic time we can find the interval corresponding to their concatenation.  Combining this with a result by Rytter on constructing small grammars to encode strings, we have given a method for preprocessing a sequence of patterns such that they can be sought quickly in a BWT-based index.  Although the preprocessing is not much faster than seaching for the patterns directly, it could be useful when we wish to search in a distributed index: we preprocess the patterns on one machine, then send them to all the machines, so that the cost of preprocessing is paid only once but the benefit is reaped for each machine.

We have also shown how the same or similar techniques can be applied to matching a pattern with wildcards in an index, obtaining a slower but more space-efficient alternative to a theorem by Lam et al.~\cite{LSTY07}; and to parallel pattern matching, showing how, given a small SLP for a pattern, we can parallelize our faster search.  We believe the techniques we have developed will prove of independent interest.

\section*{Acknowledgments}

Many thanks to Giovanni Manzini for helpful discussions.

\bibliographystyle{plain}
\bibliography{multi-pattern}
\end{document}